\documentclass[3p, 11pt]{elsarticle}
\usepackage{graphicx,amscd,amsmath,amssymb,verbatim}
\usepackage{amsfonts,epsfig}
\usepackage{mathptmx}
\usepackage{setspace}
\usepackage{epsfig}
\usepackage{url}
\usepackage{multirow}
\usepackage{color}
\usepackage{subfigure}

\begin{document}

\journal{Elsevier}

\begin{frontmatter}

\title{Towards Rock-paper-scissors patterns in the Optional Public Goods Game under random mobility}

\author{Pablo A. Valverde, Roberto da Silva, Eduardo V. Stock} 

\address{Institute of Physics, Federal University of Rio Grande do Sul,
Av. Bento Gon\c{c}alves, 9500, Porto Alegre, 91501-970, RS, Brazil
{\normalsize{E-mail:rdasilva@if.ufrgs.br}}}

\begin{abstract}

Social dilemmas concern a natural conflict between cooperation and self interests among individuals in large populations. The emergence of cooperation and its maintenance is the key for the understanding of
fundamental concepts about the evolution of species. In order to understand the mechanisms involved in 
this framework, here we study the Optional Public Good Games with focus on the effects of diffusive aspects in the emergent patterns of cyclic dominance between the strategies. Differently from other works, we showed that rock-paper-scissors (RPS) patterns occur by introducing a simple kind of random mobility in a lattice sparsely occupied. Such pattern has been revealed to be very important in the conservation of the species in ecological and social environments. The goal of this paper is to show that we do not need more elaborated schemes for construction of the neighbourhood in the game to observe RPS patterns as suggested in the literature. As an interesting additional result, in this contribution we also propose an alternative method to quantify the RPS density in a quantitative context of the game theory which becomes possible to perform a finite size scaling study. Such approach can be very interesting to be applied in other games generically.

\end{abstract}

\end{frontmatter}


\setlength{\baselineskip}{0.7cm}

\section{Introduction}

\label{Section:Introduction}

The evolution of cooperation, fairness, or pro-social behavior among
non-related individuals is one of the fundamental problems in biology and
social sciences \cite{now06}. Reciprocal altruism fails in providing a good
solution if the iterations are not repeated. Some mechanisms as punishment
can be effective, for example in the iterated ultimatum game (UG) where
players reject offers far from fifty-fifty division \cite%
{Guth1982,Silva12009,Silva22007,Almeida2014,daSilva2016419}. However this
mechanism, in the case of prisoner dilemma (PD) or even public goods games
(PGG) requires that defectors must be identified as observed in ref. \cite%
{brandt03}. Optional participation in the PGG is a simple but effective
mechanism that can avoid possible exploiters and overcome the social dilemma
~\cite{hauert02a, szabo02b}, since the cooperators and defectors can coexist
due to the abstention alternative. These works as well as many others (see
for example ~\cite{zhong13,hauert03,hauert05}) consider a dynamics with many
public games, where each one of them corresponds to a different
neighbourhood and its central node, differently from some alternative works
(not so explored in literature) that consider the dynamics of an only single
and large public goods game with interacting players (see for example \cite%
{daSilva2006610,SILVA2008}).

Therefore, the so called Optional Public Goods Game (OPGG) can provide an
useful representation of many social conflicts which the cooperation plays
an important role in the good operation of general public services.
Voluntary participation in PGG may provide a way to keep stable and
persistent levels of cooperation, without secondary mechanisms as punishment
or reward~\cite{hauert10}. In the stationary state of dynamics in OPGG the
coexistence of the three strategies: cooperators, defectors, and loners, as
well as dominance cycles of each one of these strategies in sequence, i.e.,
the so called rock-paper-scissors (RPS) regime, were reported as solutions
of the mean field replicator dynamics~\cite{hauert02b}, as well as for
simulation in different topologies~\cite{hauert03,hauert05,hauert06b}.

Other important aspects may influence the cooperation patterns in PD and PGG
and among them, one has called the attention of physicists that study
evolutionary game theory: the mobility of the players \cite{vainstein07}.
Mobility is an interesting mechanism to evaluate if a social or biological
system preserve its environment or biodiversity by considering the different
strategies in the population \cite{rei07,vainstein07}, or by simply to
change the critical rates in epidemiological systems simulated by cellular
automata \cite{silva15}. So, the investigation of RPS pattern which is an
interesting case of emergence of cooperation, deserves more attention in
OPGG and in this point we would asking for: Is the mobility an important
ingredient to influence or even preserve the RPS patterns in OPGG? If yes,
for which occupation this can happen?

In this paper we propose to study the effects of mobility in OPGG. We focus
our investigation in three different contributions:

\begin{enumerate}
\item First, we would like to answer about the connection between the
mobility and RSP patterns observed in OPPG and in other game theory
protocols;

\item In square lattices, where each site can be occupied by only one
player, we intend to explain how the occupation (density of occupied sites)
and mobility characterized by a simple diffusion parameter $p$, probability
that a player moves to a empty site, randomly chosen among nearest
neighbours, influence the RSP patterns;

\item By following this investigation, we would like to propose a parameter
to measure the density of RPS patterns in Game theory, more precisely in PD
and PGG with voluntary participation, or even in other game theory protocols;
\end{enumerate}

In this contribution we present an analysis, by means of Monte Carlo
Simulations, looking for the coexistence of the two strategies in the steady
states or in a more singular and rare case, the alternate dominance of each
single strategy (RPS patterns) in the presence of mobility. We
simultaneously analyse the effects of the multiplication factor ($r$) , the
density of mobile agents in the lattice ($\rho $) and the mobility parameter
($p$). It is important to notice that other studies considering mobility in
OPGG reveal some features to maintain the cooperation but these results do
not explore the existence of cyclic dominance of the strategies (see for
example \cite{zhong13}). More precisely, we are interested in how these
three important parameters of the game are able to modulate the emergence of
spontaneous cooperation, looking simply for coexistence of strategies or
cycles of the three possible strategies. In the next section we present the
essential points about the model and how our simulations are implemented
considering our Monte Carlo Simulations for OPGG with mobility of the
players. In section \ref{Section:Results} we present our main results.

\section{The OPGG with Monte Carlo simulations considering mobility of the
players}

\label{Section:Model}

In this paper we consider a population of $N$ players randomly distributed
over a square lattice of linear dimension $L$, so $N$ $\ \leq $ $\ L^{2}$
and the density of players is defined by $0\leq \rho $ $=$ $\frac{N}{L^{2}}%
\leq 1$. Every player interacts only with its nearest neighbours, and each
site in the lattice can contain a player or simply to be a vacancy (empty
space). So if we include the vacancy with status of state, we have a
four-state model, where each site can have four possible states: C
(cooperator), D (defector), or L (loner) if there is a player occupying the
site which represent the three possible strategies for a player and V
(vacancy) otherwise. The OPGG with Mobility evolves according to steps of
the following algorithm:

\begin{enumerate}
\item An agent $i$ is randomly chosen.

\item Each cooperator in its neighbourhood contributes to the common pool
with a unit of wealth. Defectors participate, but without contribution
(free-rider action), while loners stay out of game getting a fixed payoff $%
\sigma $, which we make equal to the unit, without loss of generality.

\item Payoffs are then calculated for the three possible strategies:

\begin{equation}
P=\left\{ 
\begin{array}{lcl}
\frac{rN_{C}}{(N_{C}+N_{D})}-1 & ; & C \\ 
&  &  \\ 
\frac{rN_{C}}{(N_{C}+N_{D})} & ; & D \\ 
&  &  \\ 
1 & ; & L%
\end{array}%
\right.
\end{equation}

where $r$ is the multiplication factor of the public good, $N_{c}$, $N_{d}$
and $N_{l}$ are the number of players with each corresponding strategy in
the local configuration (neighbourhood). If we define $s_{i}=0$ for $C$, $%
s_{i}=1$ for $D$ and $s_{i}=2$ for $L$, we can reduce this formula to:%
\begin{equation*}
P(s)=\frac{rN_{C}}{(N_{C}+N_{D})}\left[ 1-\frac{s(s-1)}{2}\right] +s-1
\end{equation*}

\item An occupied site $i$ is randomly drawn. So a site $j$ in the
neighbourhood of $i$ is also randomly drawn. If the site $j$ is a vacancy
the player localized in $i$ jumps to site $j$ with probability $p$.
Otherwise, if $P(s_{i})<P(s_{j})$ then $i$ copy the strategy of $j$. If this
condition is not satisfied, so $i$ keeps its current strategy.

\item The process is repeated $N$ times, i.e., we randomly choose $N$
players from the population, and so we replicate the process. In average
each agent has the opportunity to change its state once per Monte Carlo time
step;
\end{enumerate}

So we present time evolution densities of the three kinds of strategies, $%
\rho _{i}(t)$. In all our simulations, the initial configuration is fixed
but random, where the population is composed by 1/3 of the each one of the
three strategies, in order to observe the survival of the strategies.
Clearly, other situations should be biased to lead the system to the
emergence of cooperators. However we opted by avoiding this possible bias.
In the next section we present our main results. Basically, we define an
interesting parameter to able to describe the density of RPS patterns in the
time evolution of strategies, considering the system over different
occupations and mobilities.

\section{Results}

\label{Section:Results}

In order to investigate the collective dynamical behaviour of the system, we
measure the time evolution of fraction (or density) of individuals in the
three possible strategies, $\rho_{c}=\frac{N_{c}}{(1-\rho)L^{2}}$, $\rho_{d}=%
\frac{N_{d}}{(1-\rho)L^{2}}$ and $\rho_{l}=\frac{N_{l}}{(1-\rho)L^{2}}$,
cooperators, defectors, and loners respectively, since $(1-\rho
)L^{2}=L^{2}-N$ is the number of players in the lattice.

\begin{figure}[th]
\begin{center}
\includegraphics[width=0.7\columnwidth]{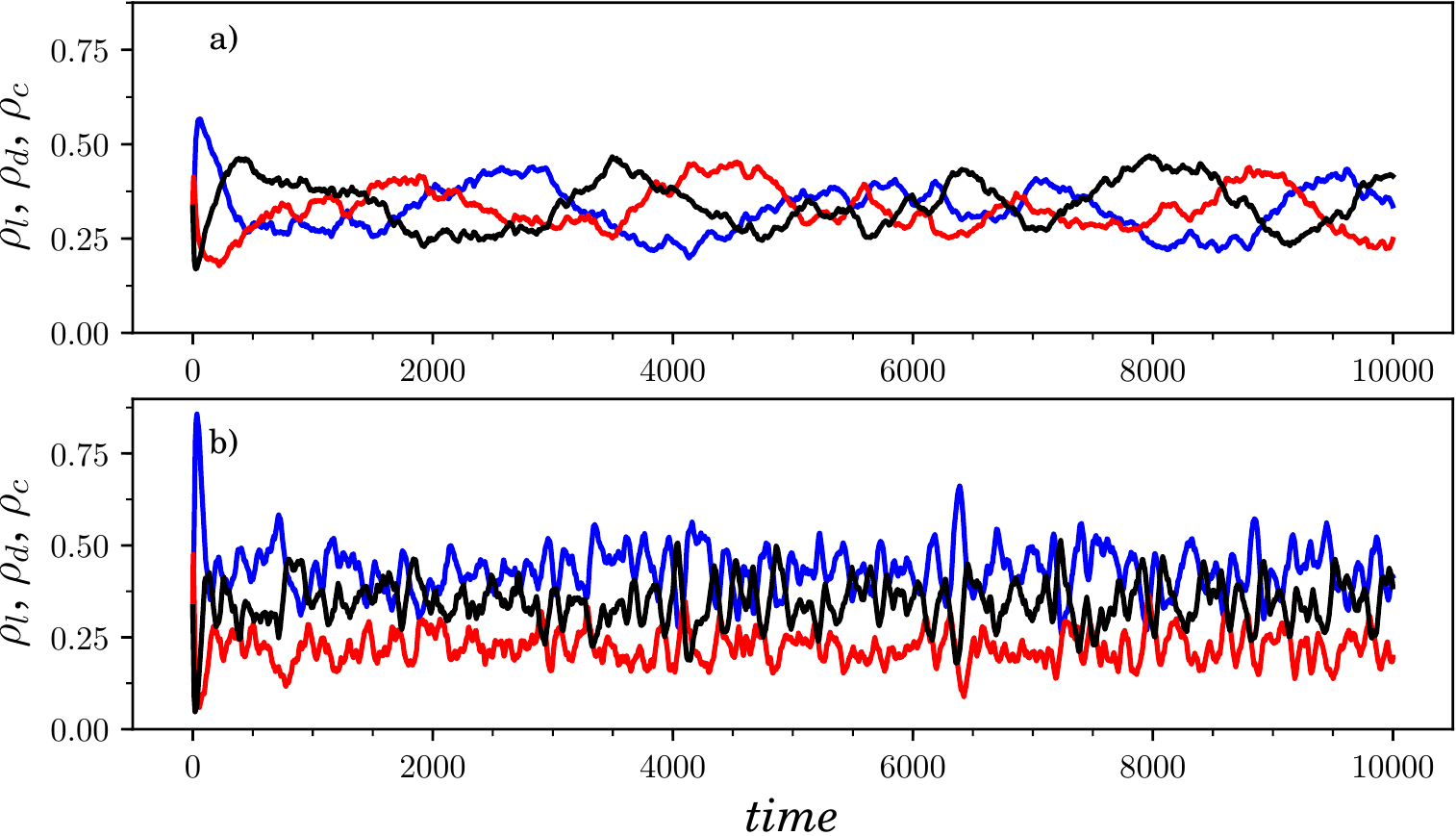}
\end{center}
\caption{ {\protect\footnotesize {\ }}Time evolution of the fraction of
cooperators (black line), defectors (red line) and loners (blue line) for $%
r=5$: a)$\ \protect\rho =0.40$, $p=0.005$ and b) $\protect\rho =0.6$, $%
p=0.005$. For that we used $L=100$, and $\protect\rho _{c}^{0}=\protect\rho %
_{d}^{0}=\protect\rho _{l}^{0}=1/3$.}
\label{fig_1}
\end{figure}

First, we start showing the time evolution of $\rho _{c}(t)$, $\rho _{d}(t)$%
, and $\rho _{l}(t)$ considering $N_{mc}=10^{4}$ Monte Carlo steps as can be
observed in Fig. \ref{fig_1}. We choose two particular situations: a) $\rho
=0.4$ and b) $\rho =0.6$, both considering $r=5$ and a low mobility: $%
p=0.005.$ As previously reported, we start from an initial concentration of $%
\rho _{c}^{0}=\rho _{d}^{0}=\rho _{l}^{0}=0.3\overline{3}$. This figure
reveals two typical patterns in our analysis: In plot (a), the system
reaches a configuration in which the three strategies coexist in a cyclic
way while in plot (b), there is no cyclic dominance among the strategies,
just coexistence where two strategies dominate one third. Now let us better
explore the effects of the vacancies and mobility of the players in the
emergence of cooperation or more precisely how this is related to the
occurrence of RPS patterns.

Here the idea is very simple: by varying the occupation and a mobility
parameters of the system we can find small fluctuations which leads to large
and sustained oscillations, so a cyclic state appears in the global
behaviour of the system, i.e., the fraction of each one must vary
cyclically. However, an important question is how to measure the density of
the occurrence of this pattern in the time evolution of these strategies in
a simple way. In this paper we suggest a natural parameter: we measure how
long is the length of all L-C-D (or C-D-L, or L-D-C...) sequences that
appear during the evolutions. So we establish the following convention: the
condition $\rho _{l}>\rho _{c,d}$ means $L$, while $\rho _{c}>\rho _{l,d}$
means $C$, and finally $\rho _{d}>\rho _{c,l}$ corresponds to the dominance
of $D$. For example, in the sequence with 40 terms: 
\begin{equation*}
\text{{\footnotesize CCLLLCCDDDLCCCCCDLCCCCDDDDCCCLLCCCCDDCCC}}
\end{equation*}%
we have 4 LCD cycles, where the density is then equal to $\alpha
=(8+7+9+8)/40=\allowbreak 0.8$, since in a general evolution we define $%
\alpha $ as 
\begin{equation*}
\alpha =\frac{\sum_{j}t_{j}}{t_{\max }}
\end{equation*}%
where\ $\left\{ t_{i}\right\} _{i=1}^{m}$ denote the set of lengths of all
sequences of LCD found in the whole series of size $t_{\max }$, where sure $%
\sum_{j}t_{j}\leq $ $t_{\max }$, which leads to $0\leq \alpha \leq 1$.

\begin{figure}[th]
\includegraphics[width=0.5\columnwidth]{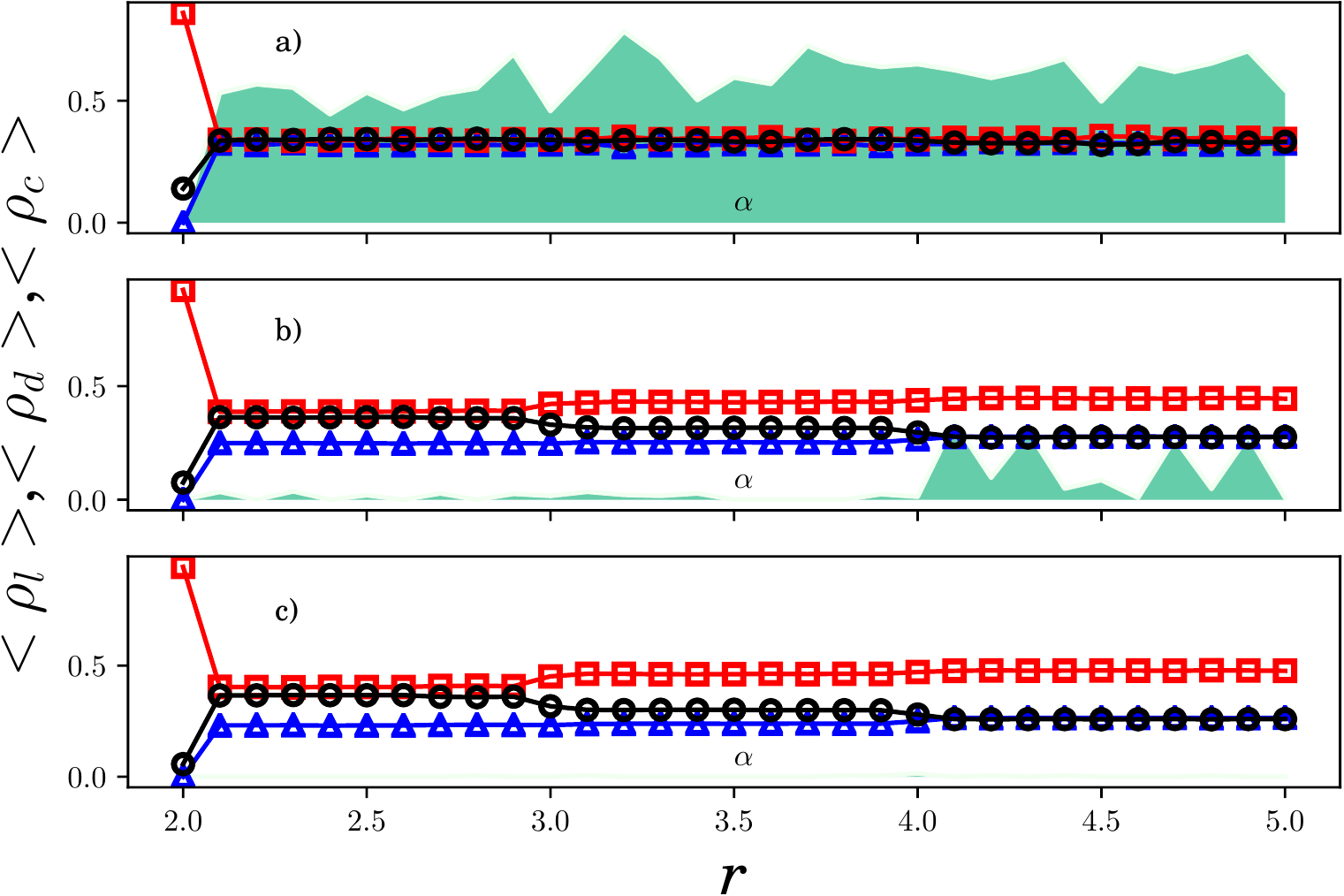} %
\includegraphics[width=0.5\columnwidth]{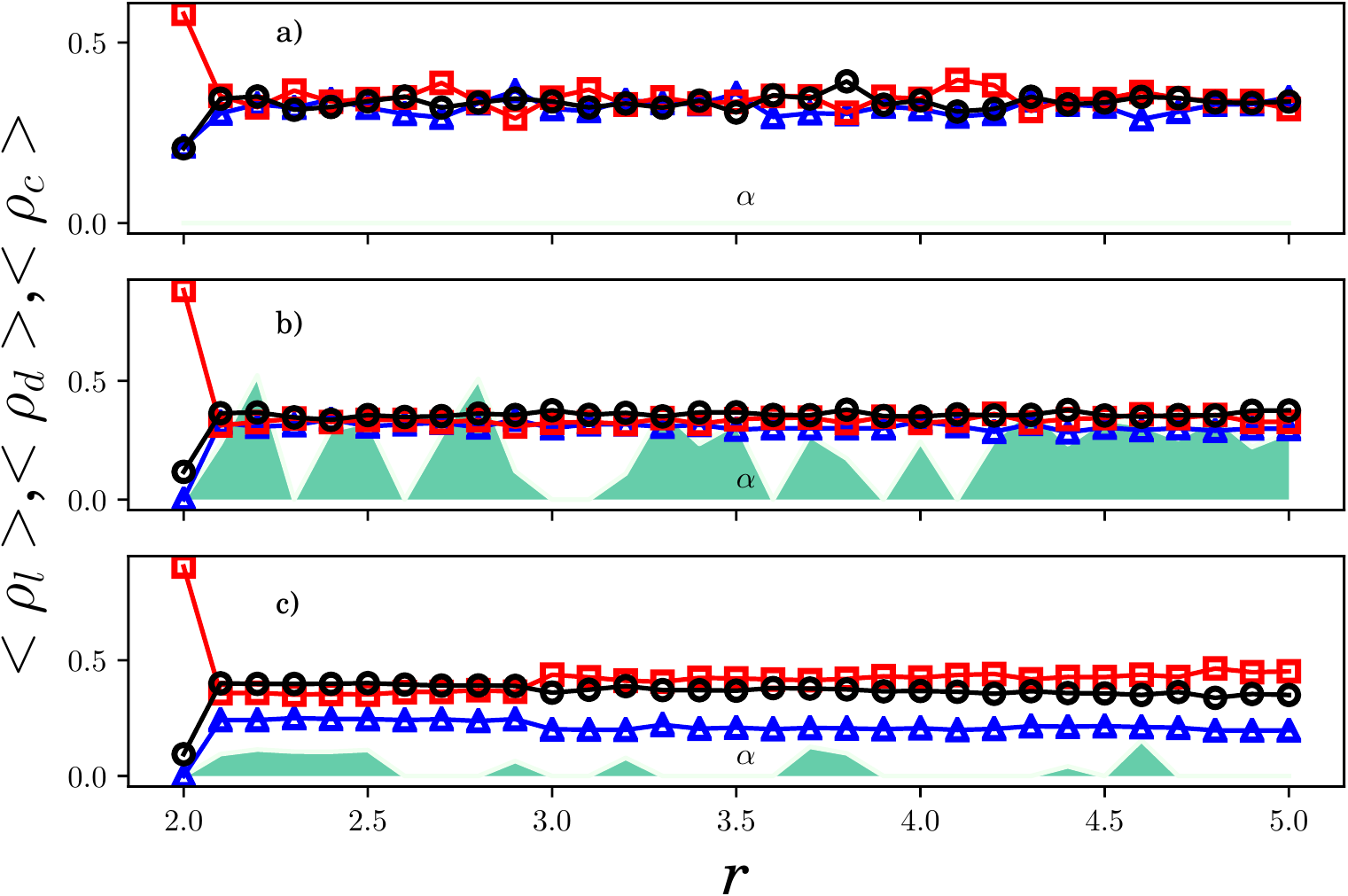}
\caption{Average frequencies of cooperators (black circles), defectors (red
squares), and loners (blue triangles) and the proportion of
rock-paper-scissors cycles ($\protect\alpha $) (green zone) as a function of
multiplication rate $r$. The three first upper plots correspond to $p=0.60$,
with the corresponding density values (a) $\protect\rho =0.1$, (b) $\protect%
\rho $ $=0.5$, (c) $\protect\rho $ $=0.6$. The three last lower plots
correspond to case $p=0.001$ for the same corresponding density values
previously used. }
\label{fig_2a}
\end{figure}

So we elaborate an interesting plot that shows $\left\langle \rho
_{l}\right\rangle $, $\left\langle \rho _{c}\right\rangle $, $\left\langle
\rho _{d}\right\rangle $ and $\alpha $ as function of $r$, the
multiplication factor of the OPPG as can be observed in Fig. \ref{fig_2a}.
The averages were computed by:

\begin{equation*}
\left\langle \rho _{i}\right\rangle =\frac{1}{t_{\max }}\sum_{t=1}^{t_{\max
}}\rho _{i}(t)
\end{equation*}%
where $i=l,c$, or $d$, and $\alpha $ were computed considering $t_{\max
}=10^{4}$ MC steps to compose the averages.

\begin{figure}[th]
\includegraphics[width=0.5\columnwidth]{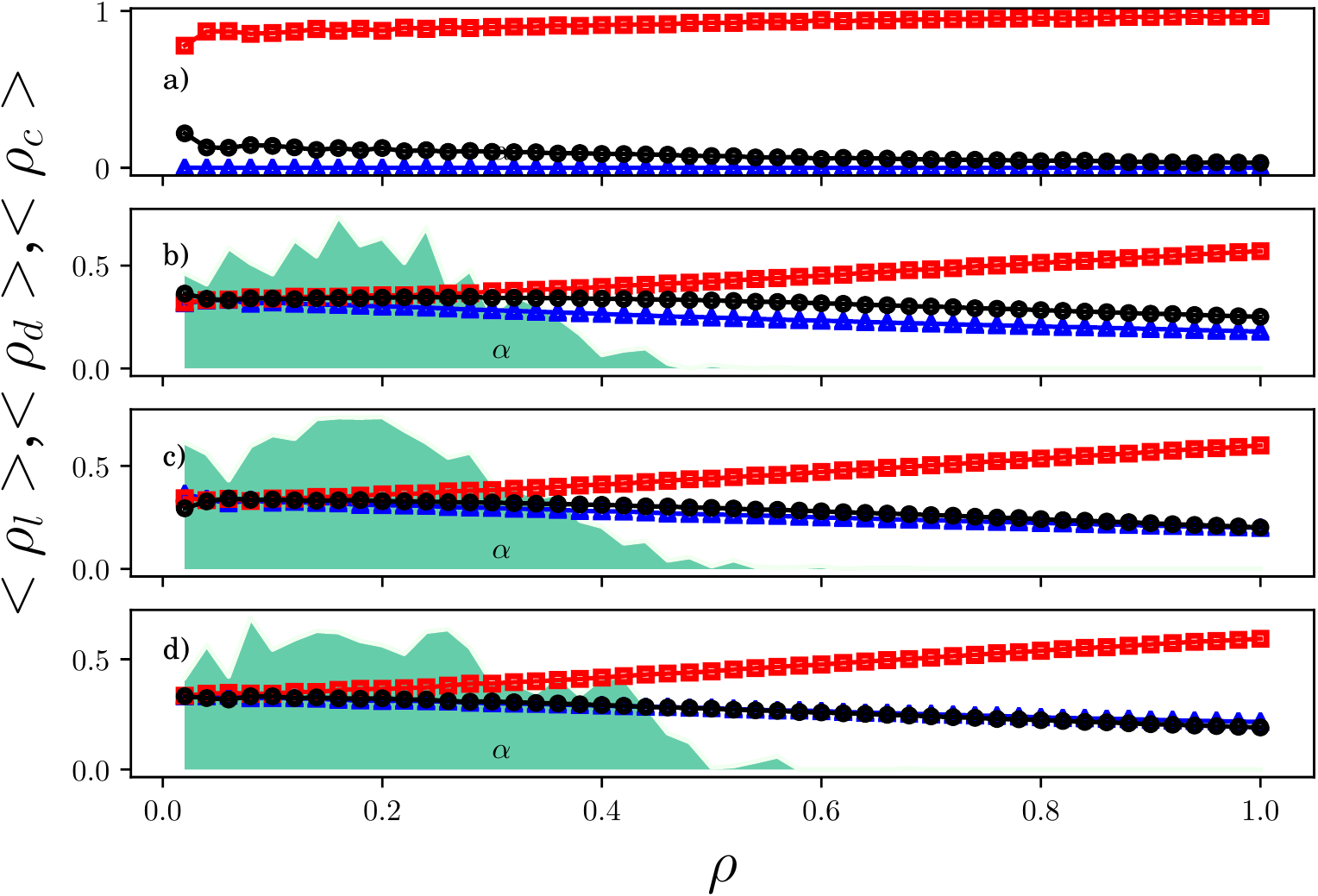}
\includegraphics[width=0.5\columnwidth]{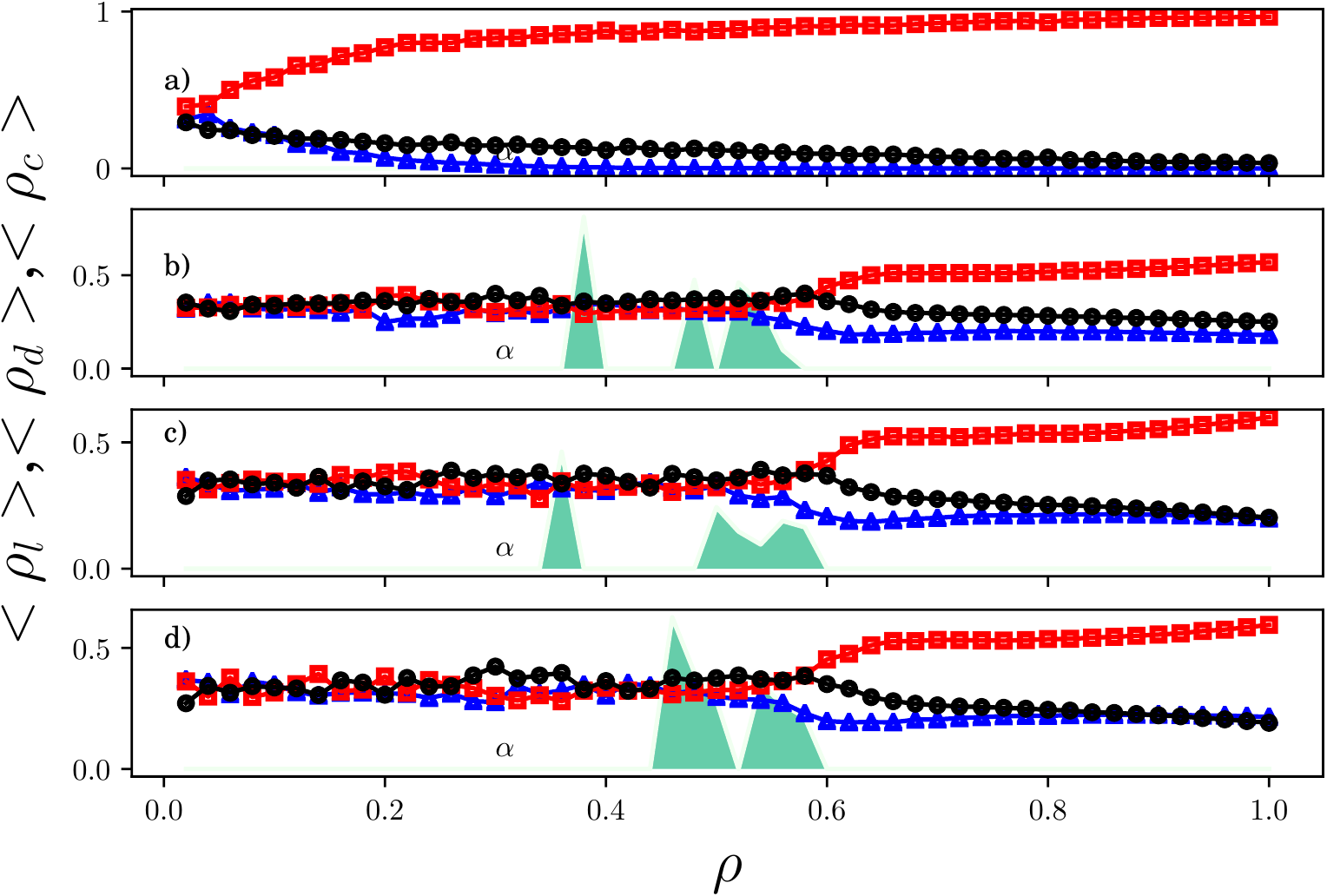}
\caption{Average frequencies of cooperators (black circles), defectors (red
squares), and loners (blue triangles) and the proportion of
rock-paper-scissors cycles ($\protect\alpha $) (green zone) as a function of
occupation $\protect\rho $. The right side plot corresponds to $p=0.60$ by
considering four different values of multiplication factor (a) $r=1$, (b) $%
r=2$, (c) $r=3$ and (d) $r=4$. The left side plot corresponds to the case $%
p=0.001$ where the same multiplication factors previously were used. }
\label{fig_2b}
\end{figure}

This figure qualitatively corroborates that there is a relation between
occupation and mobility, however this relation must be more explored. So we
perform a plot of $\left\langle \rho _{l}\right\rangle $, $\left\langle \rho
_{c}\right\rangle $, $\left\langle \rho _{d}\right\rangle $ and $\alpha $ as
function of $\rho $ as can be observed in Fig. \ref{fig_2b}.

The idea of low density and high mobility promotes RPS behaviour seems to be
qualitatively corroborated by this figure. This figure also shows that for $%
r\ <2$ the RPS patterns are not observed. However, an alternative to clear
up this point is to observe the proportion $\alpha $ to all pairs $(\rho ,p)$%
. So we elaborate color maps which show how this proportion changes in all
combinations of occupation and mobility which can be observed in Fig. \ref%
{fig_3}. We show 3 plots that correspond to the 3 different seeds
respectively, by showing the robustness of the zones obtained for the
different seeds, i.e., we can observe similar maps for the different seeds.
For the sake of simplicity, we can qualitatively separate the region in two
different regions: I which corresponds to zone with higher RPS probability
(left) and II where just coexistence of two strategies is generally observed
(right).

\begin{figure}[tbph]
\begin{center}
\includegraphics[width=0.7\linewidth,angle=0]{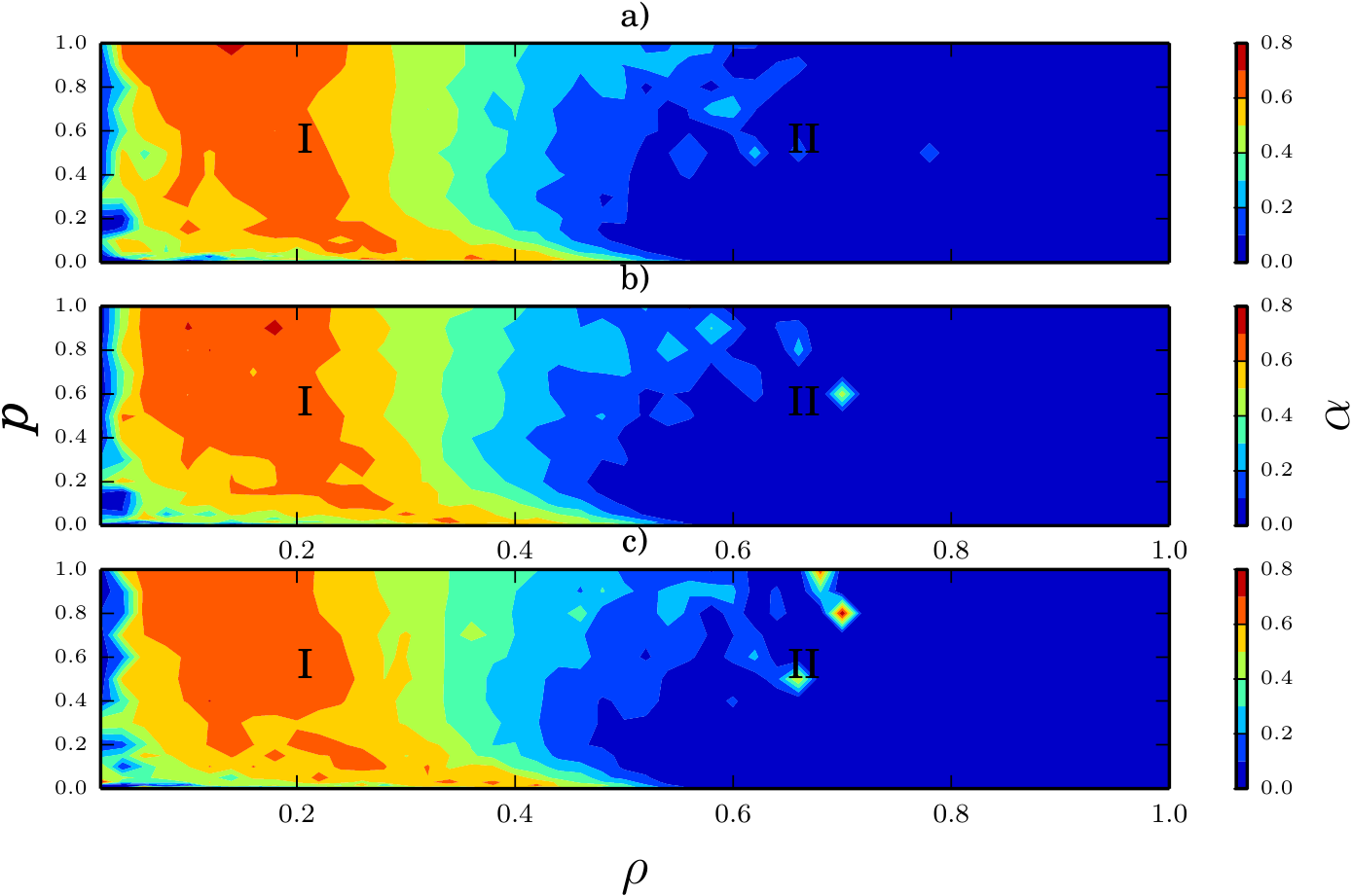}
\end{center}
\caption{\ Color Map that show the RPS density $\protect\alpha $ for each
pair ($\protect\rho ,p$) considering $r=5$. The 3 different plots correspond
to different seeds respectively used in our simulations to obtain the time
evolution strategies. }
\label{fig_3}
\end{figure}

\begin{figure}[tbph]
\begin{center}
\includegraphics[width=0.7\linewidth,angle=0]{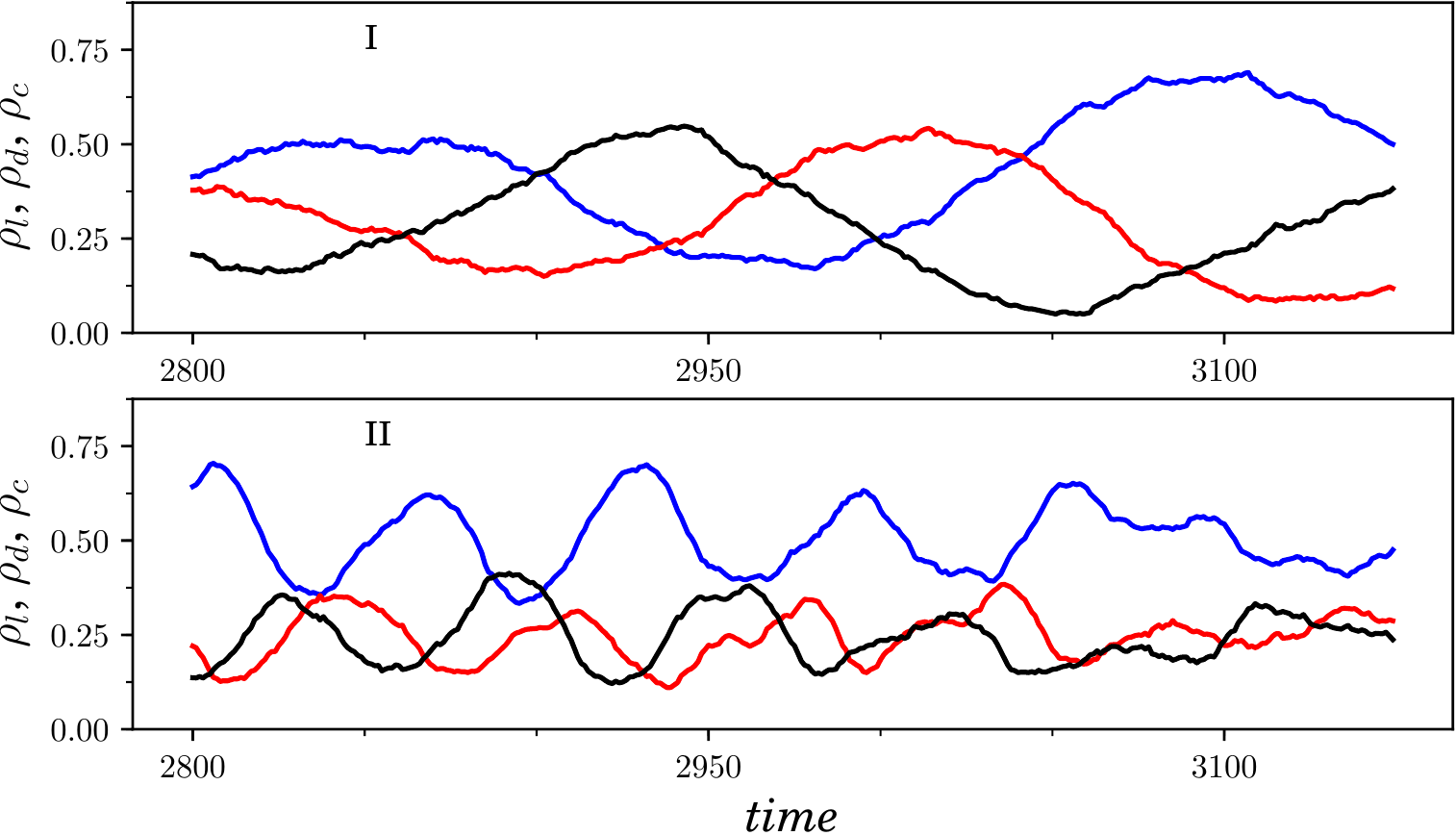} %
\includegraphics[width=0.7\linewidth,angle=0]{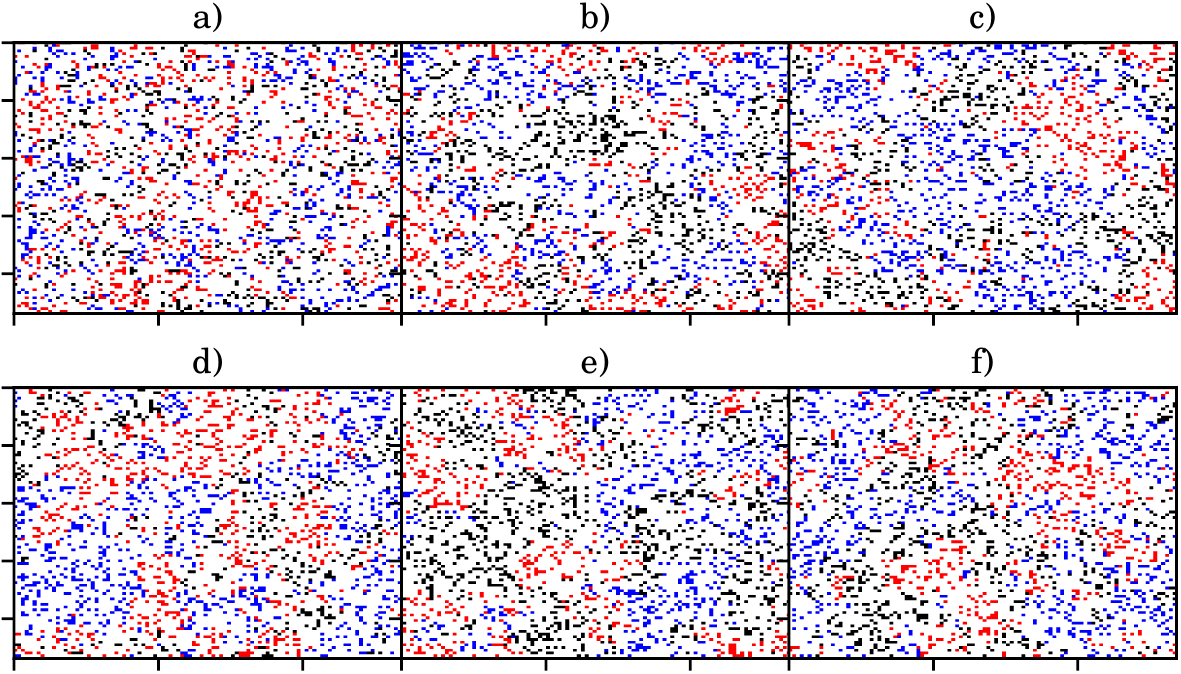} %
\includegraphics[width=0.7\linewidth,angle=0]{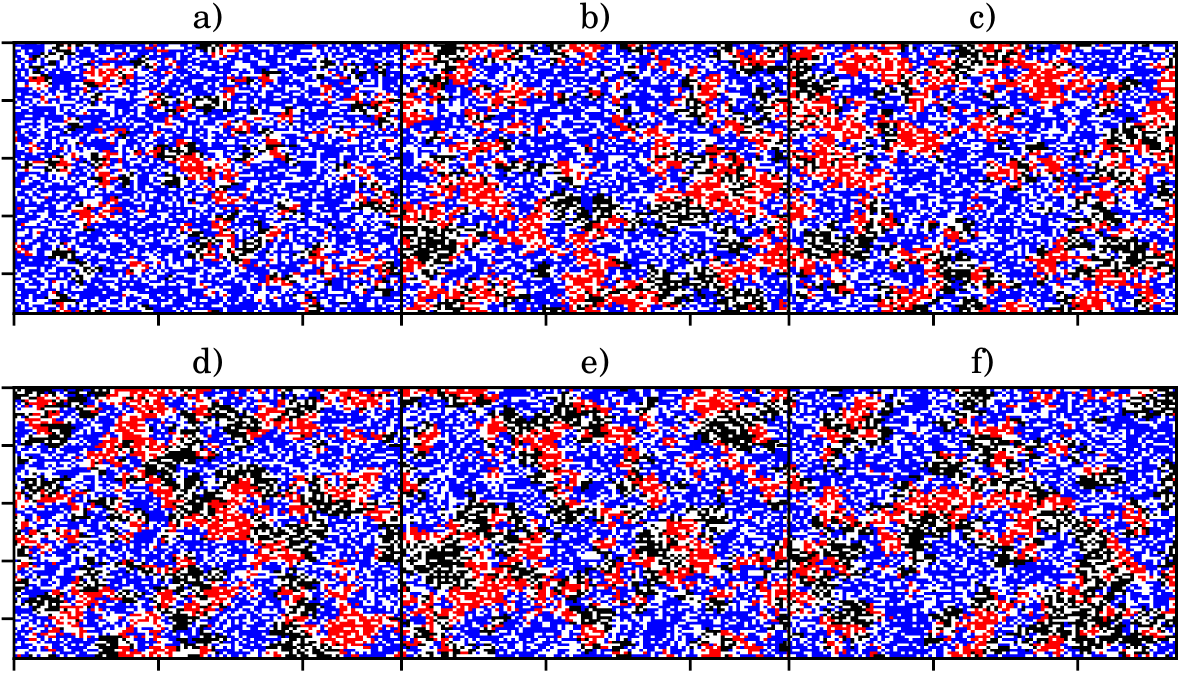}
\end{center}
\caption{\ Upper plot: The time evolutions for two different points in the
color map: I $\protect\rho =0.2$ and II: $\protect\rho =0.5$ with $p=0.5$.
Medium plot: Sequence of snapshots for the different times of evolution for
the point in I: $\protect\rho =0.2$ and $p=0.5$. Finally in the lower plot
we show the same sequence for the point in II: $\protect\rho =0.5$ and $%
p=0.5 $. }
\label{fig_4}
\end{figure}

So, let us pick up two typical points in the two different zones by
exploring their typical behaviour. For zone I we fix the parameters $\rho
=0.20$, and $p=0.5$ while in zone II we choose $\rho =0.65$, and $p=0.5$ and
so we built a figure (Fig. \ref{fig_4}) that shows the time evolution of
strategy densities for these two situations (first plot from up to bottom)
and a corresponding sequence of snapshots by showing the spatial
distribution of strategies in the lattice, for the first situation (medium
plot) and for the second situation (lower plot).

The snapshots corresponds to times a) $t=20$ b) $t=200$, c) $t=400$, d)$\
t=1000$ e)$\ t=2000$, and f) $t=5000$ MC steps. We can clearly observe the
RPS behaviour for the case chosen in region I while dominance of a strategy
is observed for the case chosen in region II. So we can conclude that with a
simple kind of mobility the RPS patterns can be observed in low occupation
and high mobility, but the pattern occurs continuously.

\begin{figure}[tbph]
\begin{center}
\includegraphics[width=0.7\linewidth,angle=0]{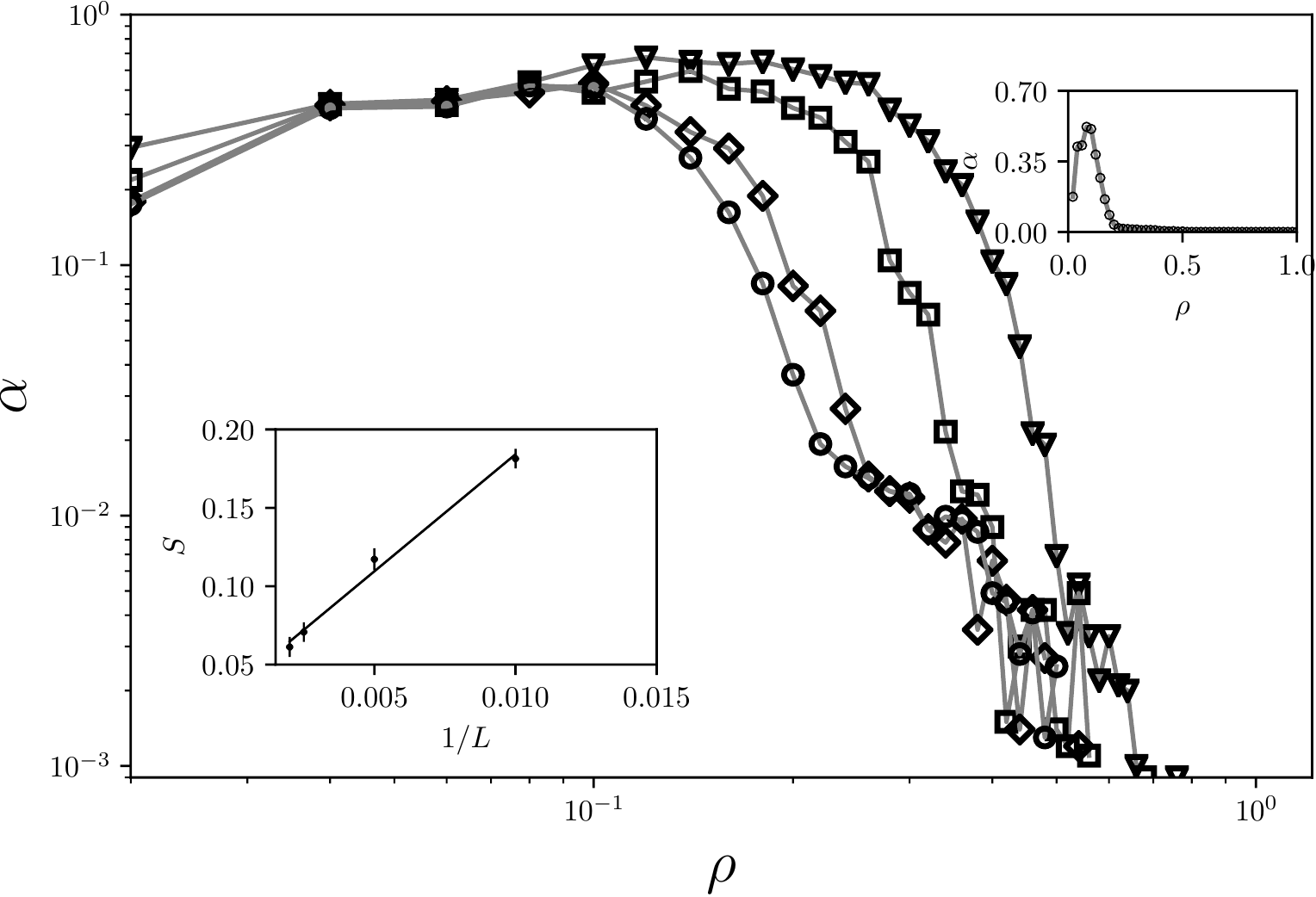}
\end{center}
\caption{{Finite size scaling effects. }Density of $RPS$ cycles $\protect%
\alpha $ as function of lattice occupation $\protect\rho $ in log-log scale.
We show plots for different size systems: $L=100$ (triangles) $L=200\ $%
(squares), $L=400\ $(losangles) and$\ L=500$ (circles). We choose the
parameters multiplicative factor $r=3.0$ and mobility $p$ $=0.5$. The upper
inset plot shows the case $L=500$ in linear scale where no fluctuations are
observed. The lower inset plot shows the RPS mass $S(L)$ as function of $L$. 
}
\label{fig_5}
\end{figure}

Finally we studied the effects of $L$ on the density of RPS. So we plot $%
\alpha (\rho )$ as function of $\rho $, for different sizes $L=100,200,400$
and $500$ which is shown in the main plot of the Fig. \ref{fig_5} in
mono-log scale. It is important to notice that we can observe strong
fluctuations for $\alpha (\rho )$ by considering different size systems. So,
in order to perform a finite size scaling analysis, we propose to use the
RPS integrated over the different $\rho $ values, which we define as RPS
mass and calculated according to: 
\begin{equation}
S(L)=\int_{0}^{1}\alpha (\rho ,L)d\rho  \label{Eq:Exact}
\end{equation}%
which is numerically obtained by Simpson's rule formulae:%
\begin{equation}
\begin{array}{lll}
S(L) & \approx & \frac{\alpha (0,L)+\alpha (1,L)}{3}\Delta \rho +\frac{2}{3}%
\Delta \rho \sum_{i=1}^{n}\alpha (\rho _{2i-1},L) \\ 
&  &  \\ 
&  & +\frac{4}{3}\Delta \rho \sum_{i=1}^{n-1}\alpha (\rho _{2i},L)%
\end{array}
\label{Eq:Simpson}
\end{equation}

We know the numerical error in approximating Eq. \ref{Eq:Exact} by Eq. \ref%
{Eq:Simpson} which is of order $O(1/n^{4})$, however we obtained the plots
of Fig. \ref{fig_5} by using five different seeds and we have the error bars
for each $\alpha (\rho _{i},L)$, represented by standard error$\ \delta
_{\alpha }(\rho _{i})$. So by discarding the numerical errors, and
considering only the statistical errors as main sources, we have the
uncertainty propagation formulae:

\begin{equation}
\delta _{S}^{2}=\frac{\Delta \rho }{3}\left[ \delta _{\alpha (0)}^{2}+\delta
_{\alpha (1)}^{2}+4\sum_{i=1}^{n}\delta _{\alpha (\rho
_{2i-1})}^{2}+16\sum_{i=1}^{n-1}\delta _{\alpha (\rho _{2i})}^{2}\right]
^{1/2}  \label{Eq.error_s}
\end{equation}

\ So we plot $S(L)$ as function of $L^{-1}$ which is shown in the lower
inset plot in \ref{fig_5} with error bars calculated according to Eq. \ref%
{Eq.error_s}. We can see the strong dependence in the size system. The upper
inset plot shows the case $L=500$ in linear scale just for observation.

We can observe that RPS mass decreases as $L$ increases according to
algebraic behaviour. So our conclusions leads to RPS occurrences in the OPGG
can be observed in no percolation regime of occupation when mobility is
incremented in the system. Our snapshots show such phenomena by showing that
mobility possibilities encounters between players that starts the RPS
process. The phenomena does not occur in highly occupied lattices and the
process is continuous, i.e., the RPS regions has a preferential region in
low occupation that changes slightly from seed to seed. Some sporadic RPS
events can be found in percolation regime ($\rho >0.6$) but are not observed
in all different color maps corresponding to the different seeds.

\section{Conclusion}

\label{Section:Conclusion}

We have performed a thorough study about emergence of cooperation in OPGG,
by showing the appearance of cycles of rock-paper-scissors patterns among
the three possible strategies in an evolutionary Darwinian dynamics under
diffusion effects in lattices with vacancies. We established color maps
which show that more probable regions of RPS patterns occur in more
sparsely lattices and intermediate mobilities while we have only coexistence
domain in more occupied lattices where the RPS patterns are missing. We also
explore the numerical relationship between the multiplicative factor $r$ of
the OPGG and the RPS patterns. We also show that RPS mass, i.e., the
integral of RPS over all possible occupations, decays as a power law as the size
system enlarges. Our work quantifies RPS patterns emergent in diluted
lattices, or only coexistence patterns, which occurs with simple mobility,
differently from other works that explore more complex diffusion to stablish
stable frequency of strategies in steady state (see for example \cite%
{zhong13}). Our work corroborates other results obtained in the context of
rock-paper-scissors games which suggest that biodiversity is promoted by
mobility (see for example \cite{rei07}).

\bibliographystyle{unsrt}
\bibliography{bibi}

\end{document}